\documentclass[reprint,amsmath,amssymb,aps]{revtex4-2}

\usepackage{graphicx}
\usepackage{dcolumn}
\usepackage{bm}

\begin{document}


\title{Quantum information theoretic measures to distinguish fermionized bosons from non-interacting fermions}
\author{Barnali Chakrabarti$^{1,2}$, Arnaldo Gammal$^2$, N D Chavda$^3$, Mantile Leslie Lekala$^4$}
\affiliation{$^1$Department of Physics, Presidency University, 86/1   College Street, Kolkata 700073, India. \\ 
$^2$ Instituto de Física, Universidade de S\~ao Paulo, CEP 05580-090, S\~ao Paulo, Brazil \\ 
$^3$ Department of Applied Physics, Faculty of Technology and Engineering,\\The Maharaja Sayajirao University of Baroda, Vadodara-390001, India \\ 
$^4$ Department of Physics, University of South Africa,
       P. O. Box 392, Pretoria 0003, South Africa.}

\date{\today}

\begin{abstract}
We study the dynamical fermionization of strongly interacting one-dimensional bosons in Tonks-Girardeau limit  by solving the time dependent many-boson Schr\"odinger equation numerically exactly. We establish  that the one-body momentum distribution approaches the ideal Fermi gas distribution at the time of dynamical fermionization. The analysis is further complemented by the measures on two-body level. Investigation on two-body momentum distribution, two-body local and non-local correlation clearly distinguish the fermionized bosons from non-interacting fermions. The magnitude of distinguishablity between the two systems is further discussed employing suitable measures of information theory, i.e.,  the well known Kullback-Leibler relative entropy and the Jensen-Shannon divergence entropy. We also observe very rich structure in the higher-body density for strongly correlated bosons whereas non-interacting fermions do not possess any higher order correlation beyond two-body. 
\end{abstract}

\maketitle

\section{Introduction} \label{intro}
The one-dimensional (1D) gas of impenetrable bosons termed as Tonks-Girardeau (TG) gas corresponds to infinitely strong repulsive bosons in the Lieb-Liniger (LL) model~\cite{Lieb:1605,Lieb:1616}, the bosons interact through a contact potential. The physical properties of 1D bosons in the TG limit coincide with those of ideal Fermi gas in the same external potential. The properties determined by the square of the wave function such as the density profile and density-density correlation become identical with those of non-interacting fermions. It leads to the interesting manifestation of fermionization when the strongly interacting bosons escape their spatial overlap~\cite{Gangardt:90,Stringari,Petrov,Girardeau,Girardeau:5}. However one-body density matrix and the momentum distribution are typically different from those of Fermi gas due to strong phase correlation originated from the bosonic statistics at the TG limit. 

The dynamics of harmonically confined TG gas exhibits more interesting scenario "dynamical fermionization" during expansion on sudden release of the trap--the momentum distribution of TG gas evolves from bosonic to fermionic character. Some aspects of dynamical evolution of the TG gas has been theoretically studied~\cite{Minguzzi,Rigol,Dominik,Yukalov}. Experimentally, optical confinement offers quasi-1D geometry and dynamical fermionization has recently been observed in the measure of asymptotic momentum distribution~\cite{Wilson}. 

Despite of well established understanding of dynamical fermionization, the topic still remains challenging as no work is reported on higher-body density and correlation measures. The existing calculations are based on the one-body level only. The aim of this work is to present some results beyond one-body measures in real as well as momentum space and to understand whether fermionized bosons can really be distinguished from ideal Fermi gas. The dynamical evolution of quasi-1D harmonically trapped strongly interacting bosons in the TG limit is investigated in this work utilizing an {\it {ab initio}} many-body technique. We solve the many bosons Schr\"odinger equation by using the multiconfigurational time dependent Hartree method for indistinguishable particles (MCTDH-X) which calculates numerically exact many-body wave function~\cite{Streltsov:2006,Streltsov:2007,Alon:2007,Alon:2008,Lode:2016,Fasshauer:2016,Lode:2020}. Measures of many-body densities and correlation functions are paramount interest to understand many-body physics.  In such strongly interacting regime we need very high orbitals to obtain the convergence in the many-body wave function. In the present work, dynamical evolution of $N=5$ fermionized bosons is studied utilizing $M=24$ orbitals in the computation of many-body wave function. 

Our work is outlined as : i) To explore the dynamical evolution of the fermionized bosons through the measures of $x$-space, $k$-space density and correlation in momentum space beyond one-body. ii) To apply information theoretic measures of relative entropy like, non-symmetric Kullback-Leibler, symmetric Kullback-Leibler and the Jensen-Shannon divergence entropy both for one-body and two-body densities. The main aim is to justify that one-body measures of density, correlation, relative entropies all are {\it{inadequate}} to understand the difference between the fermions and strongly interacting bosons which approach fermionization limit dynamically. 

Our observations are as follows:
i) As expected, dynamics of one-body densities in $x$-space are identical between strongly interacting bosons and those of non interacting fermions. ii) One-body $k$-space density exhibits that strongly interacting bosons attain fermionic character at a particular time, known as time for dynamical fermionization. iii) Measures of different relative entropies using one-body $k$-density asymptotically approach to zero at the time of fermionization. iv) Dynamics of the two-body $x$-space densities also exhibit the identical evolution for the two systems and can not extract any additional physics. v) However two-body $k$-space density of strongly interacting bosons dynamically tends to approach the fermionic two-body $k$-space density, but remains distinctly different even at the time of fermionization. vi) Measures of local and non-local two-body correlation in the momentum space quantitatively estimate the difference between the two systems. vii) Measures of relative entropies utilizing two-body density in $k$-space also clearly establish the mutual information between the fermionized bosons and fermions. viii) We additionally analyze higher-body coherence. Fermions do not possess any higher order correlation, whereas strongly interacting bosons exhibit rich structure in the three-body and four-body coherence.

All the above observations uniquely confirm that the strongly interacting bosons at TG limit dynamically acquire fermionic character but can be distinguished. The conclusion made with measures in one-body level is insufficient. Two-body correlation measures in $k$-space are the key quantities to distinguish how the ferminozed bosons differ from ideal fermions {\it{qualitatively}}, whereas information theoretic measures of relative entropies {\it{quantitatively}} distinguish the two systems.

The paper is organized in the following way. Section II briefly describes the theoretical framework. We present dynamical measures in one-, two- and higher-body levels in Section III and conclusion is made in section IV.

\section{Theoretical Framework}
In this section we introduce a brief description of the multiconfigurtaional time dependent Hartree for bosons (MCTDHB) which is used to explore the dynamics. We also introduce the basic measures which are utilized to understand the many-body dynamics.
\subsection{ Many-body Hamiltonian, wave function}
The dynamical evolution of $N$-bosons is governed by the time dependent Schr\"odinger equation 
\begin{equation}
    H|\psi(t) \rangle = i \frac{\partial}{\partial t} |\psi(t) \rangle
\end{equation}
  
The full many-body Hamiltonian is 
\begin{equation} 
\hat{H}(x_1,x_2, \dots x_N)= \sum_{i=1}^{N} \hat{h}(x_i) + \sum_{i<j=1}^{N}\hat{W}(x_i - x_j)
\label{propagation_eq}
\end{equation}
We used dimensionless unit by dividing the Hamiltonian by $\frac{\hbar^{2}}{mL^{2}}$, where $m$ is the mass of a boson and $L$ is the length scale.
Where, $h(x) = -\frac{1}{2} \frac{\partial^{2}}{\partial x^{2}} + V(x)$ is the one-body Hamiltonian. $V(x)$ is the trapping potential which is a harmonic potential 
for the present work. The interparticle interaction potential is $W(x_i-x_j) = \lambda \delta (x_i-x_j)$, $\lambda$ is the interaction strength.

In order to compute the time-evolution of the many-body Schr\"odinger equation we use the MCTDHB, which employs variational time-adaptive orbitals in the expansion of the many-boson wave function. With increase in number of orbitals it converges to the exact many-body Schr\"odinger results. In MCTDHB method, time-dependent orbitals are used in the expansion of the field operator 
\begin{equation}
    \hat{\Psi}(x) = \sum^{M}_{j=1} \hat{b}_j(t) \phi_j(x,t).
\end{equation}
Here $\{\phi_j(x,t)\}$ is a complete orthonormal set of orbitals. $\hat{b}_j$ annihilates a boson in $\phi_j$. This is in contrast with the full many-body Hamiltonian using time-independent orbitals, which require a large Fock space for convergence even for a few bosons. Introducing time-dependent orbitals, a faithful convergence can be achieved with a truncated number of orbitals. The orbitals are determined from the time-dependent variational principle ~\cite{variational1,variational2,variational3,variational4}. When $N$ bosons are distributed over $M$ time-adaptive orbitals, the time-dependent many-boson function is taken as a linear combination of time dependent permanents $\vert \bar{n};t\rangle$
 \begin{equation}
\vert \psi(t)\rangle = \sum_{\bar{n}}^{} C_{\bar{n}}(t)\vert \bar{n};t\rangle,
\label{many_body_wf}
\end{equation}
In Eq.~(4), the $\{ C_{\bar{n}} (t) \}$ are the time-dependent expansion coefficients.
The vector $\vec{n} = (n_1,n_2, \dots ,n_M)$ represents the occupation of the orbitals and $n_1 + n_2 + \dots +n_M = N$ which preserves the total number of particles. In second quantisation representation, the permanents are given as
\begin{equation}
\vert \bar{n};t\rangle = \prod_{i=1}^{M}
\left( \frac{ \left( b_{i}^{\dagger}(t) \right)^{n_{i}} } {\sqrt{n_{i}!}} \right) \vert vac \rangle.
\label{many_body_wf_2}
\end{equation}
In Eq.(4), the summation runs over all possible configurations and in the limit of $M\rightarrow \infty$, the set of permanents spans the complete Hilbert space and the expansion becomes exact. However allowing time dependent permanents, we use mush shorter expansion which leads to significant computational advantage. 

It is to be noted that both the coefficients  $\{ C_{\bar{n}} (t) \}$ and the orbitals  $\{ \phi_k(x,t) \}$ which comprise the permanents are independent parameters and are determined by the time-dependent variational principle. 
For the equations of motion governing the time evolution of $\{ C_{\bar{n}} (t) \}$ and $\{ \phi_k(x,t) \}$ we follow the variational principle based on the Lagrangian formulation~\cite{variational1}. Substituting the many-body ansatz into the functional action of the time dependent many-body Schr\"odinger equation leads to  

\begin{equation}
    \begin{split}
        S \left[ \{ C_{\bar{n}} (t) \}, \{ \phi_k(x,t)\} \right]  =   \int dt \{ \langle \Psi(t) | \hat{H} - i \frac{\partial}{\partial t} | \Psi(t) \rangle \\
        - \sum^{M}_{k,j=1} \mu_{kj}(t) \left[ \langle \phi_k | \phi_j \rangle - \delta_{kj}\right]  \}
       \end{split}
\end{equation}
Next, stationarity of the action with respect to the independent variations of $\{ C_{\bar{n}} (t) \}$ and $\{ \phi_j(x,t)\}$ are required. $\{ \mu_{kj}(t)\}$  are time-dependent Lagrange multipliers which guarantee the orthonormality of the orbitals during time propagation. \\

This is to be noted that the use of optimized time dependent orbitals leads to very fast convergence in the simulation compared to the many-body Schr\"odinger equation with time-independent orbitals,  a given degree of accuracy is reached with much shorter expansion~\cite{mctdhb_exp1,mctdhb_exp2}. We also emphasize that MCTDHB is more accurate than exact diagonalization which uses the finite basis and is not optimized. Whereas in MCTDHB, as we use a time adaptive many-body basis set, it can dynamically follow the building correlation due to inter-particle interaction~\cite{Alon:2008,Alon:2007, mctdhb_exact3,barnali_axel}. It is already established as a very efficient many-body method and used for different trap geometry and range of inter-particle interaction ~\cite{rhombik_jpb,rhombik_pra,rhombik_quantumreports, rhombik_epjd}. The MCTDHB method and the algorithm has been cast into a software package~\cite{Lin:2020,MCTDHX}.

\subsection{ Quantities of interest} 
\subsubsection{One-body and higher-order densities}
(i)The reduced one-body density matrix in coordinate space is defined as
\begin{equation}
\begin{split}
\rho^{(1)}(x_{1}^{\prime}\vert x_{1};t)=N\int_{}^{}dx_{2}\,dx_{3}...dx_{N} \, \psi^{*}(x_{1}^{\prime},x_{2},\dots,x_{N};t) \\ \psi(x_{1},x_{2},\dots,x_{N};t).
\label{onebodydensity}
\end{split}
\end{equation}
Its diagonal gives the one-body density $\rho (x,t)$ defined as
\begin{equation}
\begin{split}
\rho( x;t)=\rho^{(1)}(x_{1}^{\prime} =x \vert x_{1}=x;t)
\label{onebodydensity2}
\end{split}
\end{equation}
\vspace{1ex}
(ii) The $p$-th order reduced density matrix in coordinate space is defined by
\begin{equation}
  \begin{split}
\rho^{(p)}(x_{1}^{\prime}, \dots, x_{p}^{\prime} \vert x_{1}, \dots, x_{p};t)=\frac{N!}{(N-p)!}\int_{}^{}dx_{p+1}\,...dx_{N} \\ \psi^{*}(x_{1}^{\prime},\dots, x_{p}^{\prime},x_{p+1},\dots,x_{N};t)  \psi(x_{1},\dots,x_{p},x_{p+1}\dots,x_{N};t).
\label{pbodydensity}
\end{split}  
\end{equation}
The diagonal of $p$-body density can be represented as 
\begin{equation}
  \begin{split}
\rho^{(p)}(x_{1}, \dots, x_{p} ;t)  = \langle \psi(t) \vert \hat{\psi}^{\dagger}(x_1)\dots \hat{\psi}^{\dagger}(x_p) \hat{\psi}(x_p) \\ \dots \hat{\psi}(x_1) \vert \psi(t) \rangle
\label{pbodydensity3}
\end{split}  
\end{equation}
It provides the $p$-particle density distribution at time t. \\
To calculate the density in momentum space one needs to follow the prescription provided in the Ref~\cite{mctdhb_exact4}. Introducing $p$ $D$ dimensional Fourier transform to $r_i$ and $r_i^{\prime}$ one arrives to 
\begin{equation}
\begin{split}
\rho^{(p)}(k_1.....k_p | k_1^{\prime}.....k_p^{\prime};t)
 = \\
 \sum_{i} n_i^{(p)}(t) \alpha_i^{(p)} (k_1.....k_p,t) {\alpha_i^{(p)}}^{*} (k_1^{\prime}.....k_p^{\prime},t)
 \end{split}
 \end{equation}
 where $n_i^{(p)}(t)$ is the $i^{th}$ eigenvalue of the $p^{th} $ order reduced density matrix and $\alpha_i^{(p)} (k_1.....k_p,t)$ the corresponding eigenfunction. The eigenfunctions are known as natural $p$ functions and 
 the eigenvalues as natural occupations.
 
\subsubsection{Glauber correlation function}

The normalized $p$-th order Glauber correlation function would be the most important quantity to measure the spatial coherence. It is defined as
\begin{equation}
\begin{split}
    g^{(p)}(x_{1}^{\prime}, \dots, x_{p}^{\prime}, x_{1}, \dots, x_{p};t) = \\ \frac{\rho^{(p)}(x_{1}, \dots, x_{p} \vert x_{1}^{\prime}, \dots, x_{p}^{\prime};t)}{\sqrt{\Pi_{i=1}^{p} \rho^{(1)} (x_i \vert x_i;t) \rho^{(1)} (x_{i}^{\prime} \vert x_{i}^{\prime};t)}}
    \end{split} 
\end{equation}
The diagonal of $g^{(p)}(x_{1}^{\prime}, \dots, x_{p}^{\prime}, x_{1}, \dots, x_{p};t)$ gives a measure of $p$-th order coherence and is calculated as 
\begin{equation}
    g^{(p)}( x_{1}, \dots, x_{p};t) = \frac{\rho^{(p)}(x_{1}, \dots, x_{p} ;t)}{\Pi_{i=1}^{p} \vert \rho^{(1)} (x_i) \vert}
    \label{eq13}
\end{equation}
If $\vert g^{(p)}( x_{1}, \dots, x_{p};t) \vert =1$, the system is fully coherent, otherwise the state is only partially coherent. When $g^{(p)}( x_{1}, \dots, x_{p};t) >1$, the detection probabilities at positions $ x_{1}, \dots, x_{p}$ are correlated and $g^{(p)}( x_{1}, \dots, x_{p};t) <1$ are anti-correlated. 
Eq.~(\ref{eq13}) is further utilized to calculate the local two-particle correlation $g^{(2)}(0,0)$ and non-local two-particle correlation $g^{(2)}(0,x)$.\\

The $p^{th}$
order correlation function in momentum space is given by 
\begin{equation}
\begin{split}
    g^{(p)}(k_{1}^{\prime}, \dots, k_{p}^{\prime}, k_{1}, \dots, k_{p};t) = \\ \frac{\rho^{(p)}(k_{1}, \dots, k_{p} \vert k_{1}^{\prime}, \dots, k_{p}^{\prime};t)}{\sqrt{\Pi_{i=1}^{p} \rho^{(1)} (k_i \vert k_i;t) \rho^{(1)} (k_{i}^{\prime} \vert k_{i}^{\prime};t)}}
    \end{split} 
    \label{eq14}
\end{equation}
The diagonal of  $g^{(p)}(k_{1}^{\prime}, \dots, k_{p}^{\prime}, k_{1}, \dots, k_{p};t)$ measures the $p$-th order coherence in the momentum space. For values 
$g^{(p)}( k_{1}, \dots, k_{p};t) >1$, the detection probabilities at positions $ k_{1}, \dots, k_{p}$ are correlated and $g^{(p)}( k_{1}, \dots, k_{p};t) <1$ corresponds to loss of coherence and are anti-correlated. Eq.~(\ref{eq14}) is further utilized to calculate the local two-particle correlation $g^{(2)}(0,0)$ and non-local two-particle correlation $g^{(2)}(0,k)$.\\

\subsubsection{Information distance measures}
Shannon entropy and Fisher information are considered as the key measures to exhibit higher order characteristics in position and momentum space density distribution and has been extensively utilized in atomic systems~\cite{Robin,Aquino,Jiao,Sen,Sriraman}. 
However to measure the difference between two probability distributions over the same variable, a measure called Kullback-Leibler (KL) relative entropy is the ideal quantity~\cite{Kullback,Rao,Habibi}. Relative entropy provides mutual information between two probability distributions of the same order and defined in the same space. KL relative entropy in $x$- space is measured as 
\begin{equation}
    K(t) = \int { \rho_1^{(1)}(x,t) \ln \frac{\rho_1^{(1)}(x,t)} {\rho_{2}^{(1)}(x,t)}\;dx }
\end{equation}
Where $\rho_1^{(1)}$ and $\rho_2^{(1)}$ are the one-body densities for two different systems described in the same space and they are calculated using 
$\rho^{(1)}(x,t)$ = $\langle \psi(t) \vert \hat{\psi}^{\dagger}(x) \hat{\psi}(x) \vert \psi(t) \rangle$.

Thus, the measure of $K$ can be interpreted as a measure of deviation of $\rho_1^{(1)}$ from $\rho_2^{(1)}$ i.e, to estimate how the two systems are identical or close or far apart. Thus for any distributions $\rho_1^{(1)}$ and $\rho_2^{(1)}$, $K \geq 0$. However it is a non-symmetric measure as it depends on which distribution is considered as `reference' and which is considered as `comparison' distribution. The symmetrized Kullback distance $SK$ is defined as~\cite{Cover} 

\begin{equation}
\begin{split}
    SK(t) = \int { \rho_1^{(1)}(x,t) \ln \frac{\rho_1^{(1)}(x,t)} {\rho_{2}^{(1)}(x,t)}\;  dx }
    \\+  \int { \rho_2^{(1)}(x,t) \ln \frac{\rho_2^{(1)}(x,t)} 
    {\rho_{1}^{(1)}(x,t)}\;  dx }
    \end{split}
\end{equation}
The physical meaning of $SK$ distance is very clear. $SK$ is zero for two identical species and approach to a large value as the difference between the one-body densities of two systems $\rho_1^{(1)}$ and $\rho_2^{(1)}$ increases. 
The another symmetrized measure of relative entropy is the Jensen Shannon divergence entropy as~\cite{Lin,Lin1}

\begin{equation}
\begin{split}
    J(t) = - \int \left( \frac{ \rho_1^{(1)}(x,t) + \rho_2^{(1)}(x,t)}  {2} \right)  \times \\
    \ln \left( \frac{ \rho_1^{(1)}(x,t) + \rho_2^{(1)}(x,t)} {2} \right) dx \\
    + \frac{1} {2} \int {\rho_1^{(1)}(x,t) \ln \rho_1^{(1)}(x,t)\; dx}  \\
    +\frac{1}{2} \int {\rho_2^{(1)}(x,t) \ln \rho_2^{(1)}(x,t) \;dx}
    \end{split}
\end{equation}
Both the measures of $SK$ and $J$ are symmetrized and compare two probability distributions regardless of which distribution is considered as the `reference' and which is considered as the `comparison' distribution.
The corresponding definitions of $K$, $SK$ and $J$ measures  in $k$-space are  
\begin{equation}
    K(t) = \int { \rho_1^{(1)}(k,t) \ln \frac{\rho_1^{(1)}(k,t)} {\rho_{2}^{(1)}(k,t)} \; dk }
\end{equation}

\begin{equation}
\begin{split}
    SK(t) = \int { \rho_1^{(1)}(k,t) \ln \frac{\rho_1^{(1)}(k,t)} {\rho_{2}^{(1)}(k,t)}  dk } 
    \\ +  \int { \rho_2^{(1)}(k,t) \ln \frac{\rho_2^{(1)}(k,t)} 
    {\rho_{1}^{(1)}(k,t)}\; dk }
    \end{split}
\end{equation}

\begin{equation}
\begin{split}
    J(t) = - \int \left( \frac{ \rho_1^{(1)}(k,t) + \rho_2^{(1)}(k,t)}  {2} \right)  \times \\
    \ln \left( \frac{ \rho_1^{(1)}(k,t) + \rho_2^{(1)}(k,t)} {2} \right) dk \\
    + \frac{1} {2} \int {\rho_1^{(1)}(k,t) \ln \rho_1^{(1)}(k,t) \; dk}  \\
    +  \frac{1}{2} \int {\rho_2^{(1)}(k,t) \ln \rho_2^{(1)}(k,t) \;dk},
    \end{split}
\end{equation}
where one-body density in $k$-space $\rho^{(1)}(k,t)$ is calculated using $\rho^{(1)}(k,t)$ = $\langle \psi(t) \vert \hat{\psi}^{\dagger}(k) \hat{\psi}(k) \vert \psi(t) \rangle$.

However it is also interesting to have measures of the above three distributions using two-body density in $x$-space 
$\rho^{(2)}(x_1,x_2,t)$ = $\langle \psi(t) \vert \hat{\psi}^{\dagger}(x_1) \hat{\psi}^{\dagger}(x_2)  \hat{\psi}(x_1)  \hat{\psi}(x_2) \vert \psi(t) \rangle$.

The corresponding measures of $K$, $SK$ and $J$ in $x$ -space are defined through 
\begin{equation}
    K(t) = \int { \rho_1^{(2)}(x_1,x_2,t) \ln \frac{\rho_1^{(2)}(x_1,x_2,t)} {\rho_{2}^{(2)}(x_1,x_2,t)}\; dx_1\;dx_2 },
\end{equation}
 where $\rho_1^{(2)}$ and $\rho_2^{(2)}$ correspond to two-body densities of the two different systems but described in the same space. It will facilitate to understand whether the two systems are close in the measure of two-body perspective. The corresponding definitions of $SK$ and $J$ in $x$-space are as follows :   
\begin{equation}
\begin{split}
    SK(t) = \int { \rho_1^{(2)}(x_1,x_2, t) \ln \frac{\rho_1^{(2)}(x_1,x_2,t)} {\rho_{2}^{(2)}(x_1,x_2,t)}  dx_1 dx_2 } 
    \\+  \int { \rho_2^{(2)}(x_1,x_2,t) \ln \frac{\rho_2^{(2)}(x_1,x_2,t)} 
    {\rho_{1}^{(2)}(x_1,x_2,t)}\;  dx_1 \;dx_2 }
    \end{split}
\end{equation}

\begin{equation}
\begin{split}
    J(t) = - \int \left( \frac{ \rho_1^{(2)}(x_1,x_2,t) + \rho_2^{(2)}(x_1,x_2,t)}  {2} \right) \times
    \\ \ln \left( \frac{ \rho_1^{(2)}(x_1,x_2,t) 
    + \rho_2^{(2)}(x_1,x_2,t)} {2} \right) \;dx_1\; dx_2 
    \\+ \frac{1} {2} \int \rho_1^{(2)}(x_1,x_2,t) \ln \rho_1^{(2)}(x_1,x_2,t) \;dx_1 \;dx_2 
    \\+ \frac{1}{2} \int \rho_2^{(2)}(x_1,x_2,t) \ln \rho_2^{(2)}(x_1,x_2,t)\; dx_1\; dx_2
    \end{split}
\end{equation}

Defining two-body density in $k$-space as 
$\rho^{(2)}(k_1,k_2,t)$ = $\langle \psi(t) \vert \hat{\psi}^{\dagger}(k_1) \hat{\psi}^{\dagger}(k_2)  \hat{\psi}(k_1)  \hat{\psi}(k_2) \vert \psi(t) \rangle$ we can have the measures of $S$, $SK$ and $J$ in momentum space from two-body perspective. They are calculated as 
\begin{equation}
    K(t) = \int { \rho_1^{(2)}(k_1,k_2,t) \ln \frac{\rho_1^{(2)}(k_1,k_2,t)} {\rho_{2}^{(2)}(k_1,k_2,t)} \; dk_1\;dk_2 }
\end{equation}

\begin{equation}
\begin{split}
    SK(t) = \int { \rho_1^{(2)}(k_1,k_2, t) \ln \frac{\rho_1^{(2)}(k_1,k_2,t)} {\rho_{2}^{(2)}(k_1,k_2,t)}  dk_1 dk_2 }
    \\ +  \int { \rho_2^{(2)}(k_1,k_2,t) \ln \frac{\rho_2^{(2)}(k_1,k_2,t)} 
    {\rho_{1}^{(2)}(k_1,k_2,t)} \; dk_1 \;dk_2 }
    \end{split}
\end{equation}

\begin{equation}
\begin{split}
    J(t) = - \int \left( \frac{ \rho_1^{(2)}(k_1,k_2,t) + \rho_2^{(2)}(k_1,k_2,t)}  {2} \right) \times
    \\ \ln \left( \frac{ \rho_1^{(2)}(k_1,k_2,t) 
    + \rho_2^{(2)}(k_1,k_2,t)} {2} \right) 
    \;dk_1 \;dk_2 
   \\ + \frac{1} {2} \int \rho_1^{(2)}(k_1,k_2,t) \ln \rho_1^{(2)}(k_1,k_2,t)\; dk_1\; dk_2 
    \\+ \frac{1}{2} \int \rho_2^{(2)}(k_1,k_2,t) \ln \rho_2^{(2)}(k_1,k_2,t)\; dk_1\;dk_2
    \end{split}
\end{equation}

\begin{figure}[tbh]
\centering
\includegraphics[width=0.4\textwidth, angle=0]{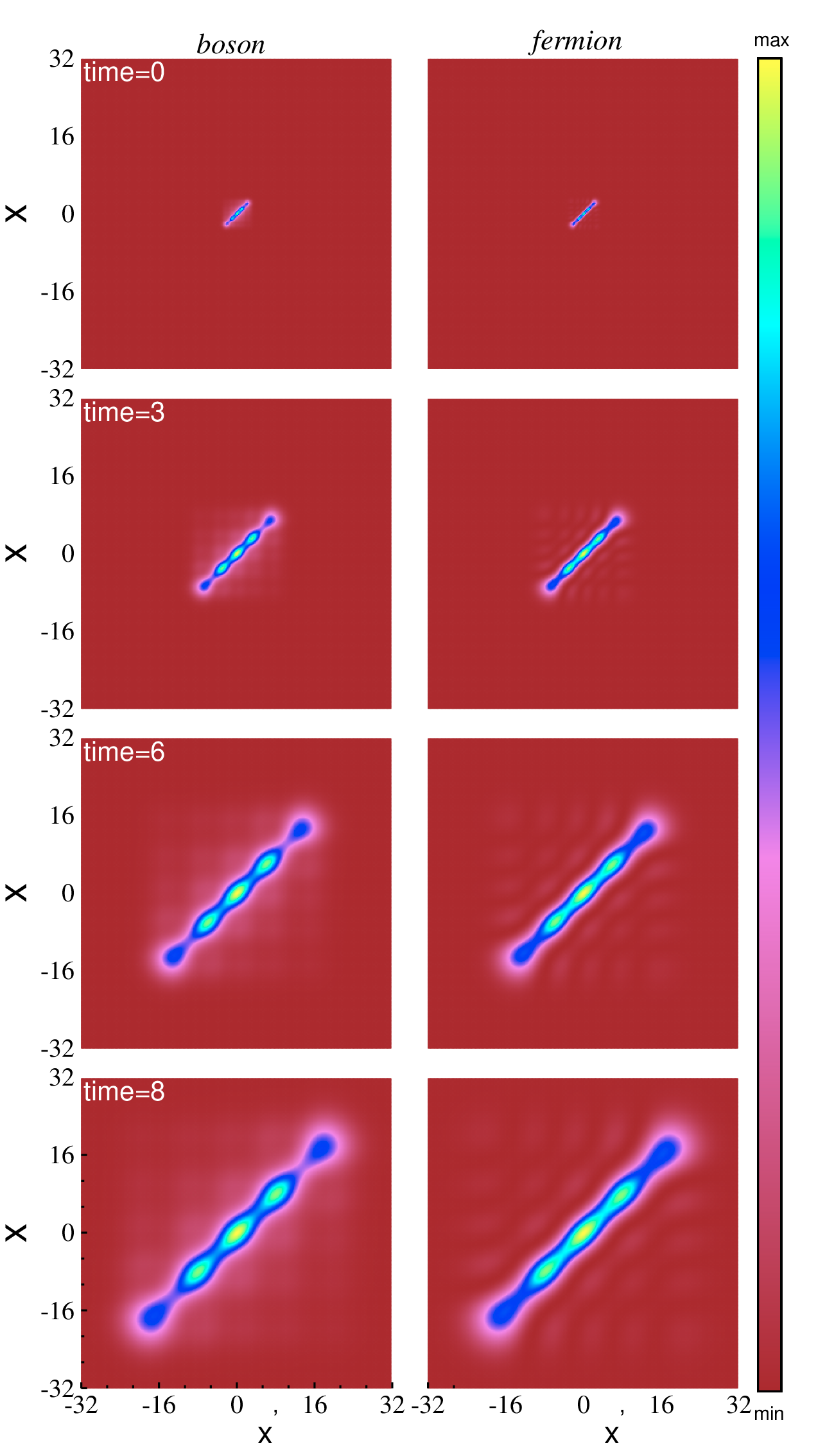}
\caption{Snapshots of the time evolution of reduced one-body density in the $x$-space ($\rho^{(1)} (x, x^{\prime})$) during the expansion of $N=5$ strongly interacting bosons with interaction strength $\lambda=25$ are presented (left panel). Computation is done with $M=24$ orbitals. Comparison is made for the dynamics of non interacting fermions in the right panel, computation is made with $M=5$ orbitals. The dynamical evolution of the two systems is identical in $x$-space.}
\label{fig1}
\end{figure}

\begin{figure}[tbh]
\centering
\includegraphics[width=0.71\textwidth, angle=-90]{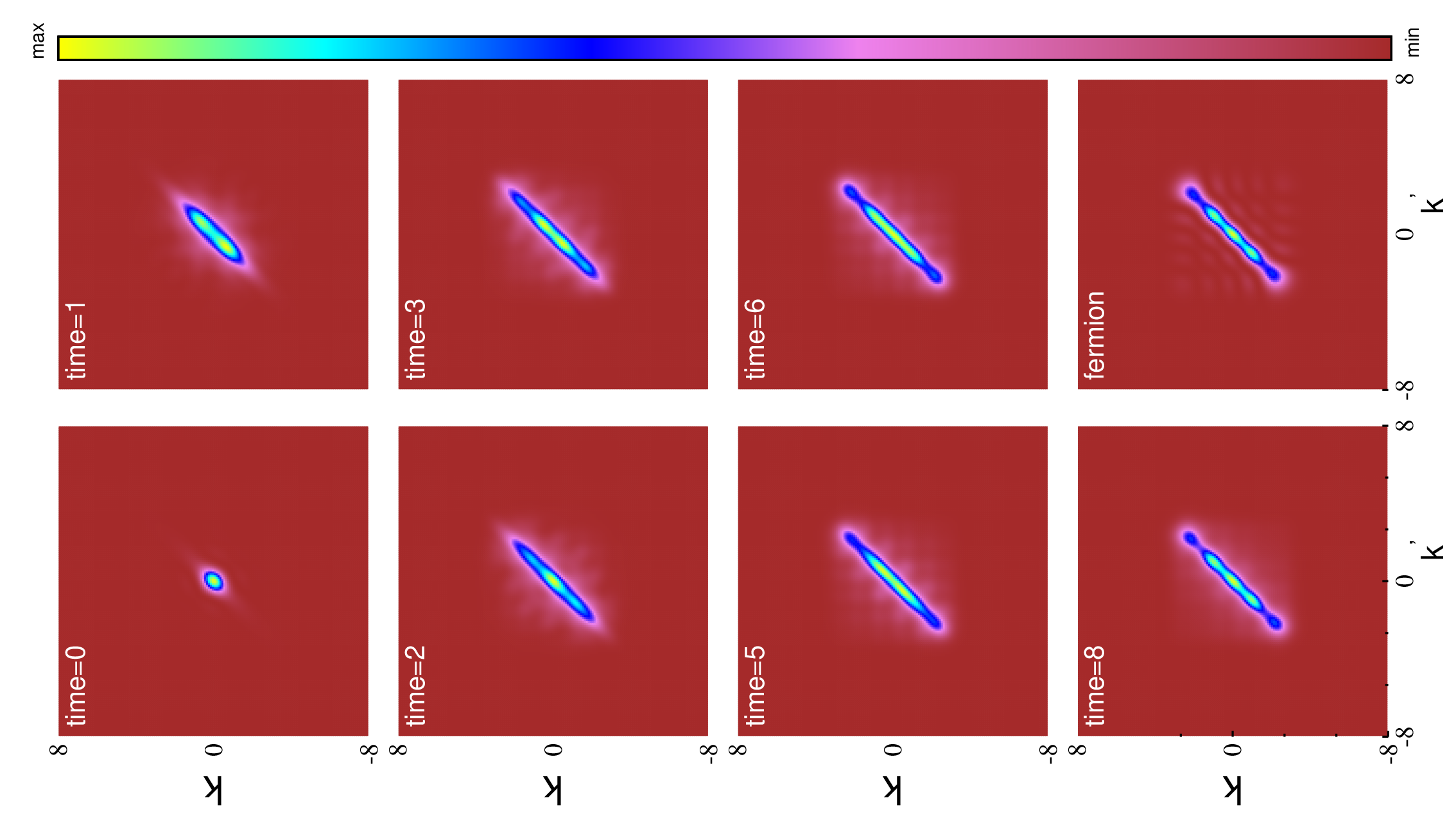}
\caption{Snapshots of the time evolution of reduced one-body density in the $k$-space ($\rho^{(1)} (k, k^{\prime})$) during the expansion of $N=5$ strongly interacting bosons with interaction strength $\lambda=25$ are presented for different times. Computation is done with $M=24$ orbitals. Comparison is made with fermionic momentum density distribution (last panel in the right side), computation is made with $M=5$ orbitals. At time $t=8.0$, strongly interacting bosons attain the fermionic density distribution.}
\label{fig2}
\end{figure}

\section{Results of expansion dynamics}
\subsection{Measures on one-body level}
We choose $N=5$ strongly interacting bosons with $\lambda =25$, in the harmonic oscillator trap $V(x) = \frac{1}{2} x^{2}$. We use MCTDHB for bosons and for the noninteracting fermions we use MCTDHF (multiconfigurational time-dependent Hartree for fermions) implemented in the MCTDH-X software package~\cite{MCTDHX}. The computation is done with $M=24$ orbitals for strongly interacting bosons, the convergence is assured as the last orbital occupation is insignificant. For noninteracting fermions, computation is done with $M=5$ orbitals, each orbital has exactly equal contribution of $20 \%$ population. The initial state is prepared at the ground state in the harmonic oscillator and the 1D expansion is studied on sudden removal of the trap. We monitor the dynamical evolution in one and higher-order many-body densities.\\
Fig.1 depicts a comparison of dynamical evolution of reduced one-body density in $x$- space $\rho^{(1)}(x, x^{\prime})$ between the strongly interacting bosons (left panel) and non interacting fermions (right panel), exhibiting identical expansion dynamics. Initially at $(t=0)$, the density is clustered at the center due to the harmonic trap, but becomes flatter and broader with increase in time. The density gradually acquires modulations and the number of humps tries to saturate to the number of bosons. At $t=8$, we observe emergence of five distinct humps. The hump at the centre is the brightest where the potential was initially zero, the two outer humps are less pronounced due to the initial non-zero confining potential and the two outermost humps are least pronounced which are away from the center. It is clearly seen that both strongly interacting bosons and non-interacting fermions exhibit identical expansion dynamics in $x$-space. We additionally note that the density's maxima in the TG regime are distinct but not isolated, the five bright spots are interconnected.  
\begin{figure}[tbh]
\centering
\includegraphics[width=0.4\textwidth, angle=0]{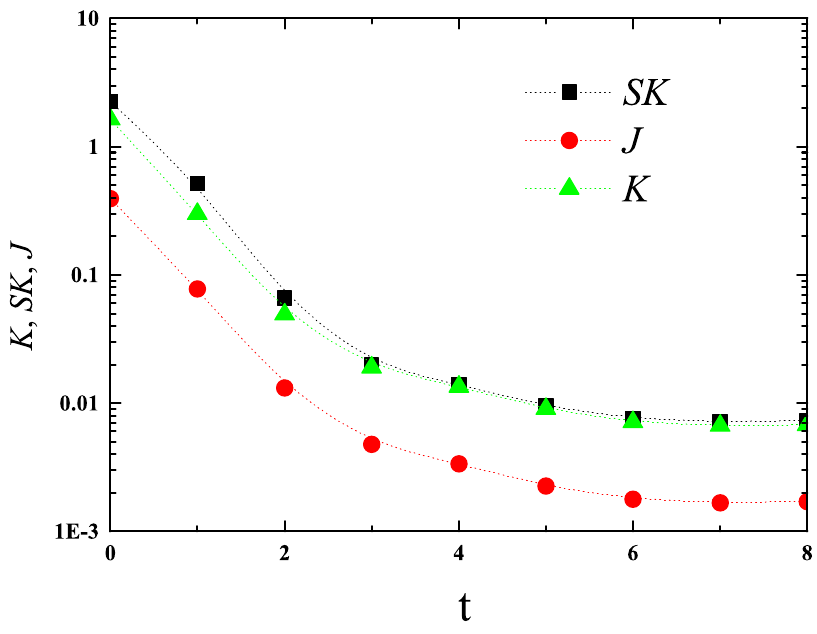}
\caption{Dynamics of relative entropy measures $K$, $SK$ and $J$ in momentum space as determined by the equations (18), (19) and (20) respectively. The corresponding one-body momentum density is utilized to evaluate the entropy measures. All the measures exhibit how the strongly interacting bosons asymptotically approach fermionization limit at time $t=8.0$ when all the relative entropy measures reach close to zero value.}
\label{fig3}
\end{figure}
Next we study the expansion dynamics through reduced one-body density in $k$-space, i.e $\rho^{(1)}(k,k^{\prime})$ which evaluates the time for dynamical fermionization. Fig.2 describes the momentum distribution of the expanding gas with five bosons at different time ($t$= 0,1,2,3,5,6,8). Initially at $t=0$, the density is in cluster form at the center of the trap, with time it gradually develops the peaks, the peaks become prominent and finally become identical with the fermionic density distribution (last panel on the right side) at $t=8.0$. We observe dynamical fermionization occurs as the momentum distribution approaches ideal Fermi distribution at time $t=8.0$. It is not possible to distinguish the fermionized bosons and the ideal fermions from the measure of one-body $k$-density. We conclude $t=8$ is the time for dynamical fermionization for the present system. 

We calculate the measures of relative entropies $K, SK$ and $J$ in the $x$-space using Eqs.(15),(16) and (17) respectively. However as shown in Fig.1, as the fermionized bosons and noninteracting fermions exhibit identical one-body expansion dynamics in real space, the corresponding relative entropy measures are identically zero all throughout the dynamical evolution. The corresponding measures in momentum space as determined by the Eqs. (18), (19) and (20) are plotted in Fig.3. All the measures have the same trend and conclude the same physics. $K$ and $SK$ almost overlap whereas $J$ lies below.  We find, initially the values are significant which infer that the two systems, strongly interacting  bosons and fermions are distinctly different. With time, all the three measures gradually decrease which signify that the strongly interacting bosons gradually attain the fermionic characteristics. Finally at the time of fermionization ($t=8.0$), they asymptotically reach close to zero which signifies that the density of the two systems become identical. 

All the measures based on the dynamics of one-body density support the well accepted fact that at the time of fermionization the strongly correlated bosons asymptotically achieve fermionization. The one-body momentum density of the two systems become indistinguishable.  However it needs to examine whether the same conclusion can be made from the measures using two-body densities.

\subsection{ Measures on two-body level}

We discuss dynamics of the reduced two-body density in $x$-space $\rho^{(2)}(x,x^{\prime})$ in Fig.4. Comparison between strongly interacting bosons and non-interacting fermions are presented at the same time scale as of one-body density [Fig.1]. Left panel corresponds to the strongly interacting bosons and the right panel corresponds to the non-interacting fermions. Initially, the atoms are clustered near the center ($x=x^{\prime}=0$) for both the cases of interacting bosons and non-interacting fermions. With time, $\rho^{(2)}$ gradually spreads out to the off-diagonal $(x \neq x^{\prime})$. The diagonal $(x\simeq x^{\prime})$ is gradually depleted and a so-called "correlation hole" forms on the diagonal. Thus the bosons gradually overcome their spatial overlap; the probability of finding two bosons at the same position tends towards zero. With time, the 'correlation hole' spreads and finally converges in the fermionization limit at time $t=8.0$. Off-diagonal spreading also converge in the fermionization time. Similar to the one-body density, the maxima which are formed across the off-diagonal are distinct but not isolated. We infer that the correlation hole along the diagonal and the confined spreading at the time of fermionization are the unique signatures of the
two-body density of a dynamically evolved fermionized state. The dynamics exhibited by the non-interacting fermions is exactly identical to that of interacting bosons.

\begin{figure}[tbh]
\centering
\includegraphics[width=0.4\textwidth, angle=0]{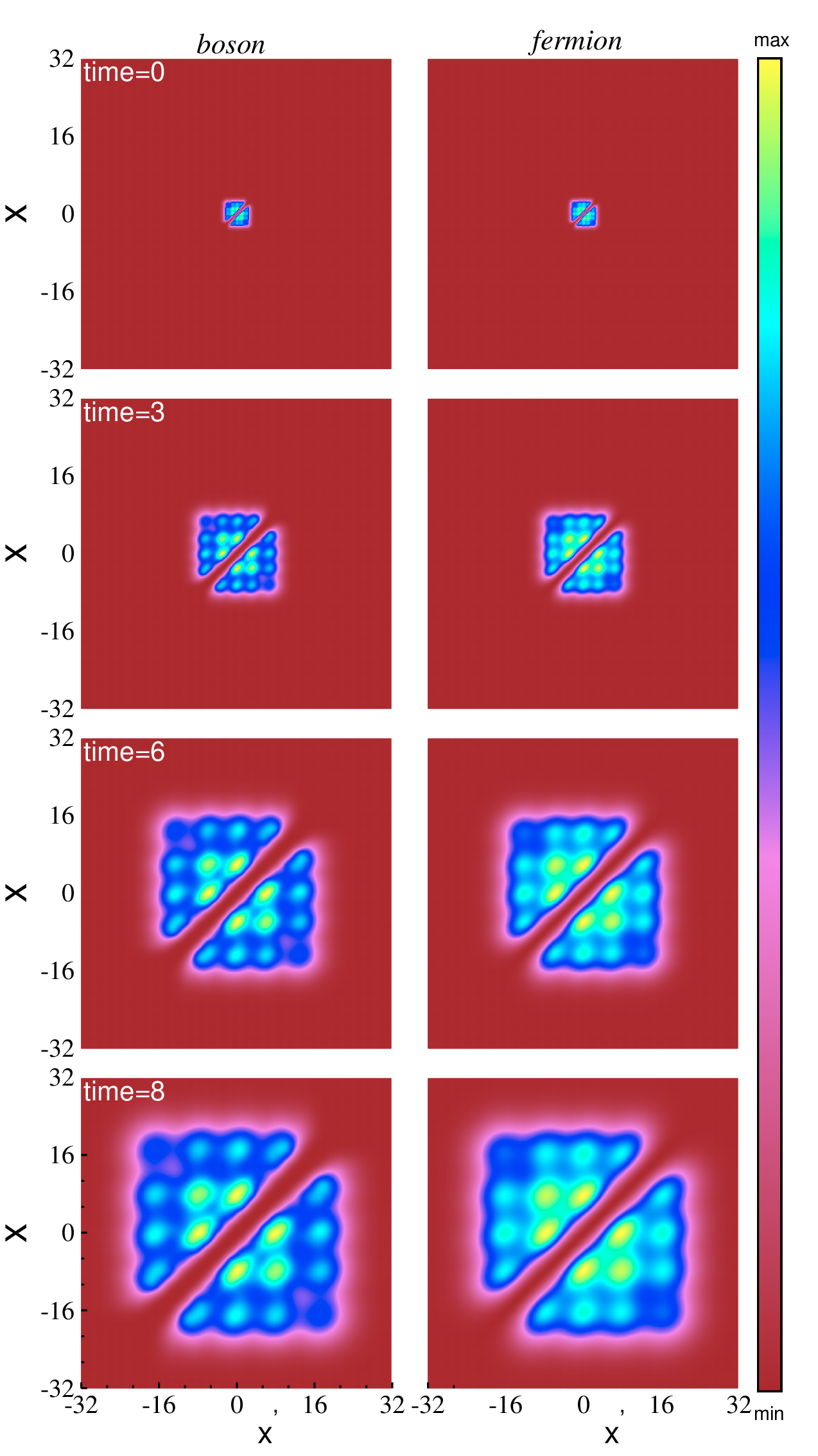}
\caption{Snapshots of the time evolution of reduced two-body density in $x$-space ($\rho^{(2)} (x, x^{\prime})$) during the expansion of $N=5$ strongly interacting bosons with interaction strength $\lambda=25$ are presented (left panel). Computation is done with $M=24$ orbitals. Comparison is made with dynamics of expansion for non- interacting fermions with $M=5$ orbitals (right panel). Identical evolution of the two systems is observed. }
\label{fig4}
\end{figure}

\begin{figure}[tbh]
\centering
\includegraphics[width=0.71\textwidth, angle=-90]{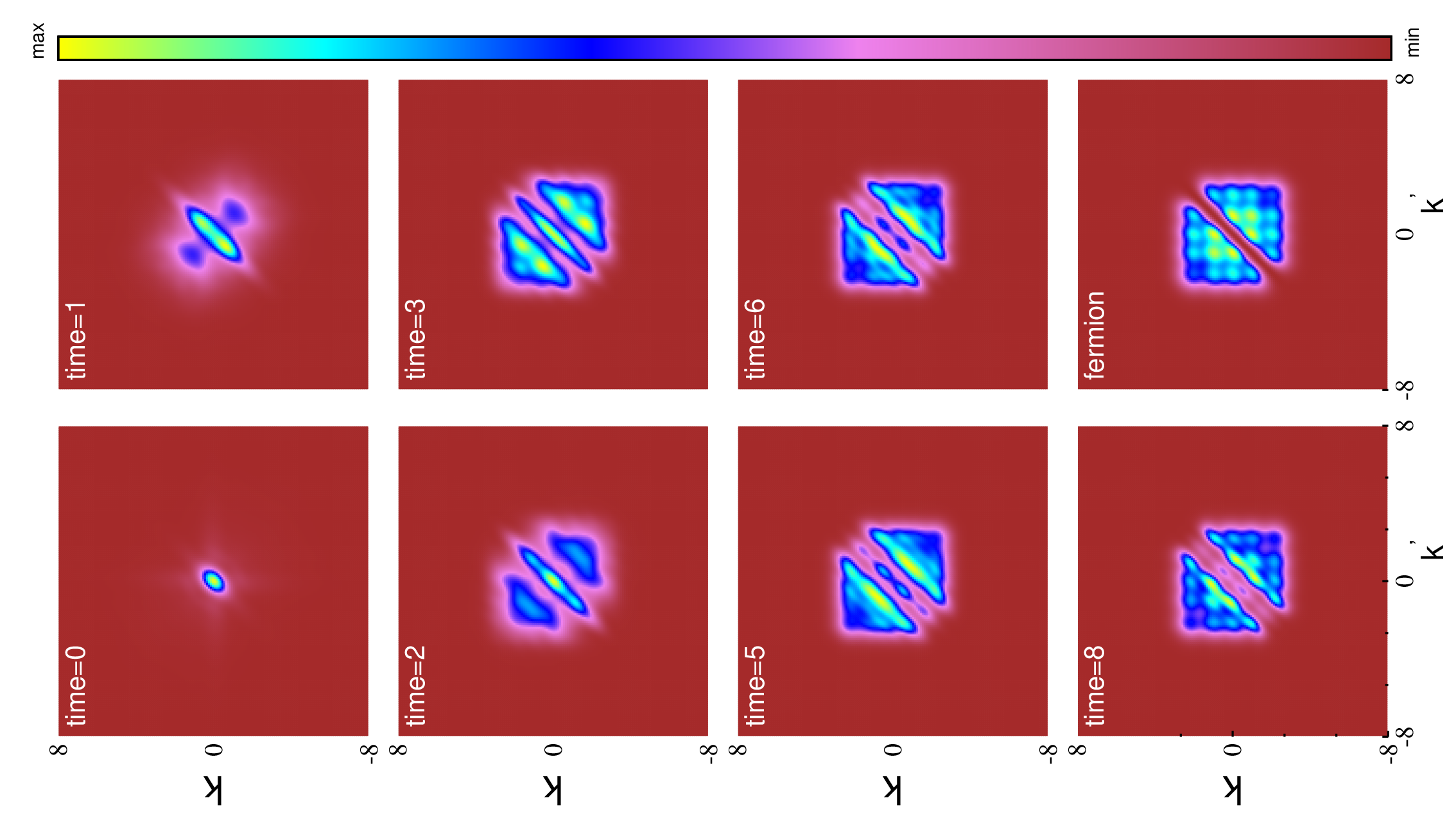}
\caption{Snapshots of the time evolution of the reduced two-body momentum density ($\rho^{(2)} (k, k^{\prime})$) during the expansion of $N=5$ strongly interacting bosons with interaction strength $\lambda=25$ and orbitals $M=25$ are presented for different times. The corresponding fermionic distribution computed with $M=5$ is presented at the last panel on right side. The strongly interacting bosons which was initially at the center of the trap, gradually develops very weak correlation hole along the diagonal. However at the time of dynamical fermionization, the correlation width and intensity strongly differ from the distinct correlation hole in the fermionic distribution. }
\label{fig5}
\end{figure}

In Fig. 5, we plot the reduced two-body density in $k$-space $\rho^{(2)}(k, k^{\prime})$  for the same time points as chosen for Fig.2. Initially at time $t=0$, we observe the density in the clustered state at the center. With time  $\rho^{(2)}$ spreads out to the off-diagonal and the diagonal is depleted. However no distinct correlation hole is created, the strongly interacting bosons try to make a diagonal gap, but some internal structure along the diagonal exist. $\rho^{(2)}(k,k^{\prime})$ remains localized near $k \leq 3.5$, because of the finite energy. The well defined bright spot indicates the localization of the bosons in momentum space. With further time evolution the internal structure along the diagonal is greatly reduced but not completely extinguished like ideal fermionic two-body reduced momentum distribution. 

To understand qualitatively how close the strongly interacting bosons feature the fermionic properties in the two-body level, we utilize the information distance measures. The measures utilizing two-body densities in $x$-space from Eqs. (21), (22) and (23) result to zero as the fermionized bosons and noninteracting fermions exhibit indistinguishable expansion dynamics as shown in Fig.4. The corresponding measures utilizing two-body momentum density from Eqs (24), (25) and (26) are plotted in Fig.6. The initial large value in all the relative entropy measures clearly distinguish that strongly interacting boson's density is completely different from that of fermionic two-body density. With increase in time, the entropy measures fall which signifies that the two-body density of bosons try to achieve fermionic two-body density. At the time of fermionization ($t=8.0$), all the measures settle to a significant minimum value but far above zero leading to infer that dynamically fermionized bosons have distinguishable density from that of non-interacting fermions in the two-body level. It signifies that although at the time of fermionization, one-body density of strongly interacting bosons becomes identical to that of ideal fermions, but their corresponding two-body density differs.  
We conclude that the bosons in TG limit is close to fermions in 'two-body perspective analysis', but not identical as concluded from 'one-body perspective' analysis.  

\begin{figure}[tbh]
\centering
\includegraphics[width=0.4\textwidth, angle=0]{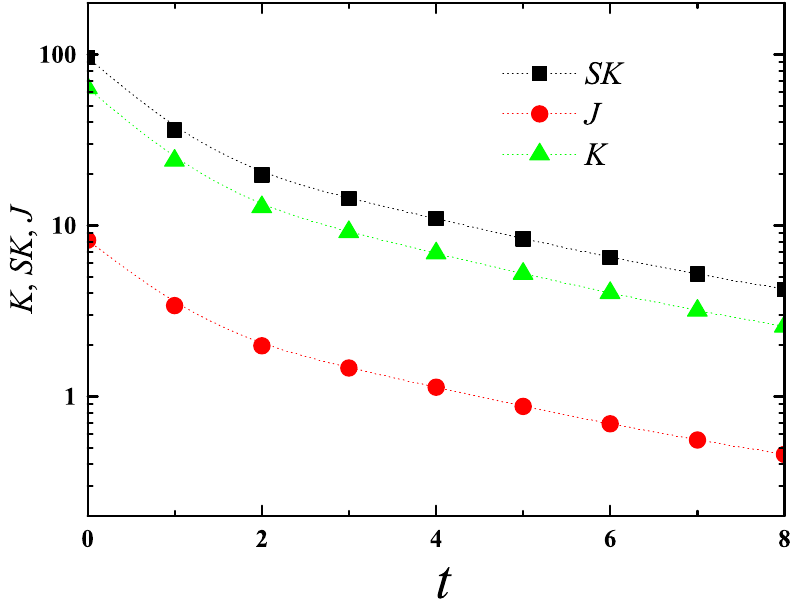}
\caption{Dynamics of relative entropy measures between the strongly interacting bosons and the ideal fermions using two-body momentum density in the equations (24)  to (26). The initial large value clearly signify how the two systems are quite far from each other. With time all the three measures gradually decreases with time as the strongly interacting bosons try to acquire fermionic distribution. However unlike the observation made in Fig. 3, $K$, $SK$ and $J$ settle to a significant minimum value.}
\label{fig6}
\end{figure}

\begin{figure}[tbh]
\centering
\includegraphics[width=0.4\textwidth, angle=0]{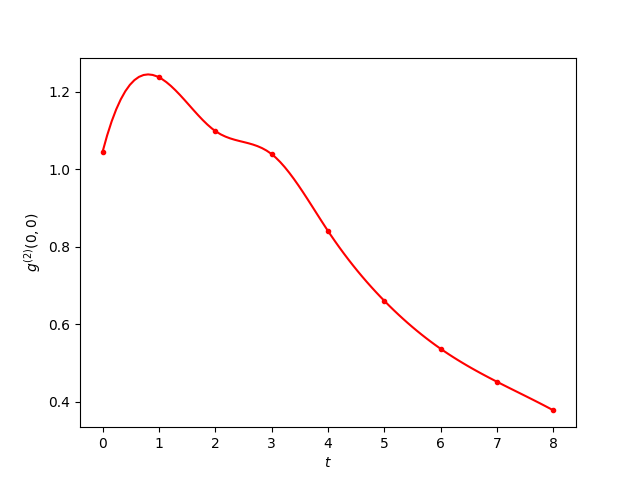}
\caption{The time evolution of the local two-body correlation $g^{(2)}(0,0)$ for the strongly interacting bosons in momentum space. $g^{(2)}(0,0)$ = $0$ corresponds to fermionization limit which infers existence of distinct correlation hole; two-body correlation is completely extinguished. Strongly interacting bosons initially have strong local correlation, it dies with time but does not reach to zero at the time of fermionization.}
\label{fig7}
\end{figure}

\begin{figure}[tbh]
\centering
\includegraphics[width=0.4\textwidth, angle=0]{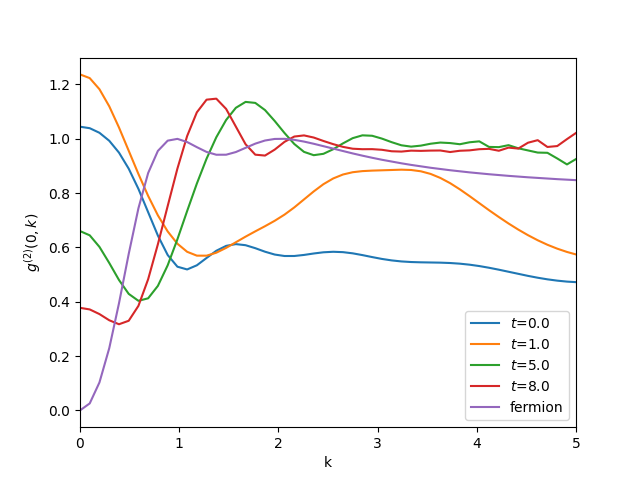}
\caption{The time evolution of two-body non-local correlation $g^{(2)}(0,k)$ for the strongly interacting bosons.  $g^{(2)}(0,k)$ $\rightarrow$ $1$ implies the 'no correlation' limit which is achieved at the characteristic time of fermionization. We also observe some peculiar oscillation around the 'no correlation' zone for the strongly interacting bosons. }
\label{fig8}
\end{figure}

To estimate further the difference between the strongly interacting bosons and the fermions we calculate the local two body correlation function  both in $x$ and $k$-space. From Fig. 4, it is clearly seen that the two-body correlation is completely extinguished at $x=x^{\prime}=0$ all throughout the dynamics both for the fermionized bosons and non-interacting fermions, it leads to $g^{(2)}(x=0, x^{\prime}=0)$ $=0$ as calculated from the Eq. (13).  However in the $k$-space, the local two-body correlation $g^{(2)}(k=0, k^{\prime}=0)$ would contain significant quantitative information about the close proximity of the two systems in the entire dynamics. From the Eq. (14) we calculate $g^{(2)}(k=0, k^{\prime}=0)$ for fermions and fermionized bosons and plot it in Fig. 7. It is expected that for fermions $g^{(2)}(0,0)$ is uniquely zero which is the typical feature of fermionization. For strongly interacting bosons it starts from $g^{(2)}(0,0) =1.0$, and then after a small increase, it smoothly decreases. However at the time of fermionization, it settles to $g^{(2)} =0.4$. It indicates that for strongly interacting bosons the local two-body correlation is not completely extinguished, which helps to identify strongly interacting bosons from non-interacting fermions. 
The corresponding two-body non-local correlation $g^{(2)}(0,k)$ is plotted in Fig. 8.
Strongly interacting bosons reach to the 'no correlation' limit with some initial characteristic oscillation. However the fermions attains the 'no correlation'  limit sharply and does not exhibit any resolvable dynamical structure. 

\subsection{Higher-order coherence}
Fig.9 and Fig. 10 depict the higher-order densities $\rho^{(p)} (p=3,4)$ in $x$ and $k$-space for the strongly interacting bosons; the non-interacting fermions do not exhibit any higher-order correlation beyond two-body. The $\rho^{(p)}$ provides the probability of detecting particles at $x$ and $x^{\prime}$ when the remaining $p-2$ particles are fixed at some reference positions. For fermionized bosons, the diagonal of the high-
order $p$-body densities $\rho^{(p)}(x, x)$ vanishes as  the  so-called correlation hole results from the strong interaction strength that mimics the Pauli principle,
preventing to find two bosons at the same position. The maxima exhibit well defined peaked structure which indicates the localization of
the atoms in position space. The maxima along the anti-diagonal $x=-x^{\prime}$ infers that the
bosons maximize the distance between each other. Whereas the maxima along the sub-diagonal infers that the bosons minimize the
potential energy. The additional correlation holes appear at the fixed values of the remaining $p$ $-$ $2$ coordinates of $\rho^{(p)}$, preventing to find other bosons at these positions.

In Fig.9, we plot three-body density $\rho^{(3)} (x, x^{\prime})$ (keeping the third particle at the fixed reference point $x_3=0$) at the left column and four-body density $\rho^{(4)} (x, x^{\prime})$ (keeping the third and the fourth particle at the fixed reference points $x_3=0$ and $x_4=0.47$) at the right column. Initially both three-body and four-body density are localized at the center. With time, the maxima is developed along the anti-diagonal as well as along the sub-diagonal.  Additional correlation holes also appear at the fixed reference points as probability of finding other bosons at these positions is zero. The three-body correlation structure exhibit further delocalization as expected and finally converge at the time of dynamical fermionization. The corresponding three-body and four-body densities in the momentum space are plotted in Fig. 10. Like Fig.9, it also exhibits rich dynamical structure.\\

\begin{figure}[tbh]
\centering
\includegraphics[width=0.4\textwidth, angle=0]{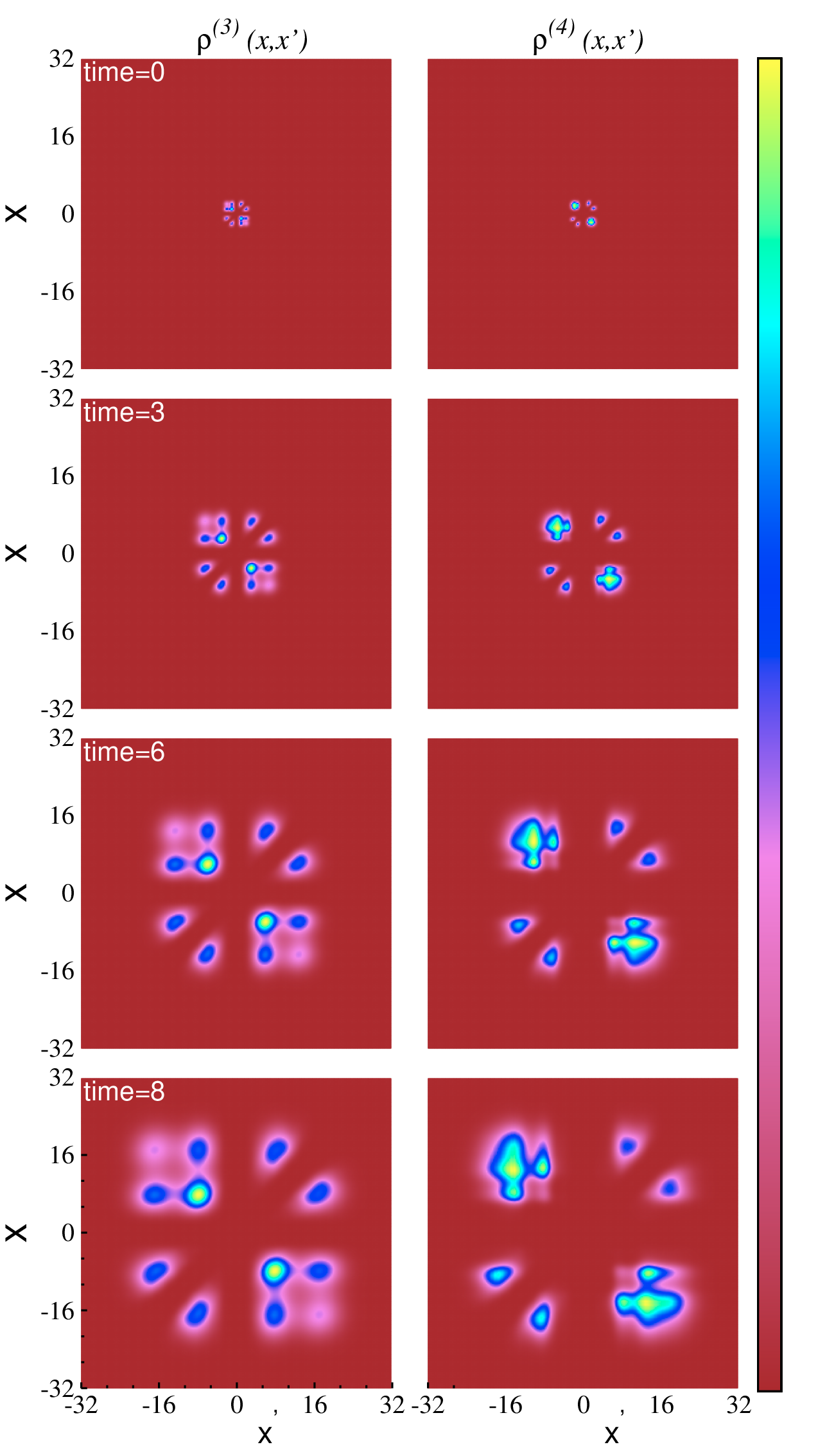}
\caption{Snapshots of the dynamics of three-body density $\rho^{(3)}(x,x^{\prime})$ (left column) and $\rho^{(4)} (x, x^{\prime})$ (right column) for $N=5$ strongly interacting bosons. See the text for details. }
\label{fig9}
\end{figure}

\begin{figure}[tbh]
\centering
\includegraphics[width=0.4\textwidth, angle=0]{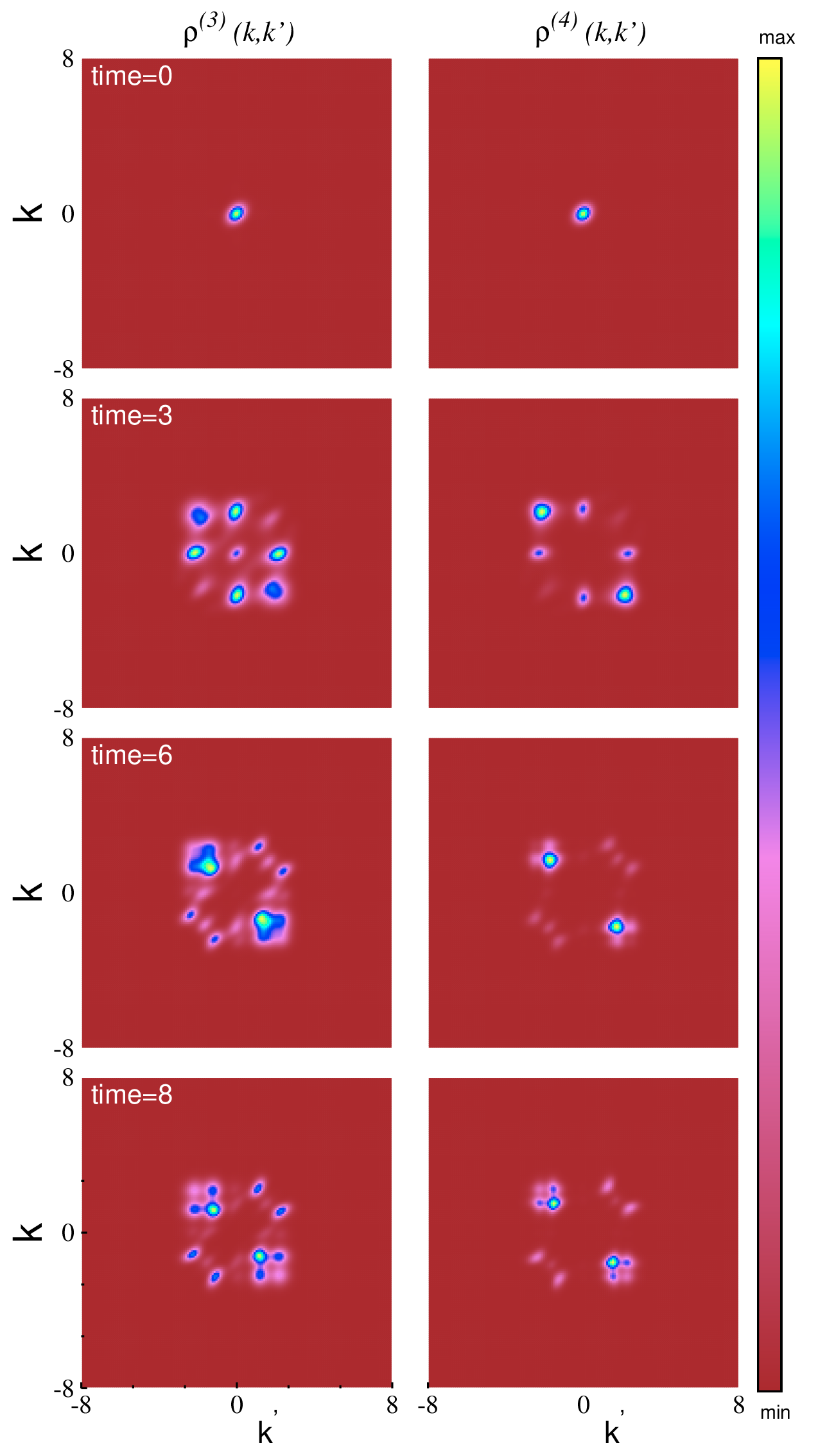}
\caption{Snapshots of the dynamics of three-body density $\rho^{(3)}(k,k^{\prime})$ (left column) and $\rho^{(4)} (k, k^{\prime})$ for $N=5$ strongly interacting bosons. See the text for details.}
\label{fig10}
\end{figure}

\section{Conclusion}
For the strongly interacting TG limit, the wave function of the interacting bosons in one dimension maps to that of non interacting fermions. The properties of the two systems which are determined by the amplitude of the wave function are identical. Whereas the measurement in the momentum space exhibits the dynamical fermionization. This conclusion is established on the basis of one-body analysis. In this work we extend the measure to higher order densities and correlation function using numerically exact solution of the time dependent many-boson Schr\"odinger equation. We establish all the existing physics and additionally include the measures of three different kinds of relative entropies which utilize the one-body in the momentum space. We clearly find that all the measures on one-body level unambiguously establish the onset of fermionization in the dynamical expansion of strongly interacting bosons. However the measures beyond one-body level is able to distinguish the strongly interacting bosons from noninteracting fermions in the dynamics. To understand that how the two systems reach to close proximity at the onset of fermionization we extend all the measures in two-body level. The two-body momentum distribution clearly distinguish the two systems in the dynamical evolution. The measures of relative entropies using two-body momentum density also quantitatively exhibit the distinction between the two systems. The two-body local and non local correlations also conclude the same difference. Additionally we find that the interacting bosons exhibit very rich structure in the higher body densities whereas for the fermions three- and higher body correlation are ideally zero. 

\textit{Acknowledgements --}
This work was supported by the FAPESP grant Process No. 2023/06550-4. We are thankful to Rhombik Roy for helpful discussion.

\bibliography{roy}
\end{document}